\documentclass[aps,prl,reprint,floatfix,amsmath,amssymb, groupaddress, showkeys]{revtex4-2}

\usepackage{bbm}
\usepackage{graphicx}
\usepackage{hyperref}
\usepackage{bm}
\usepackage{printlen}
\usepackage{units}
\renewcommand{\vec}[1]{\mathbf{#1}}

\hypersetup{hidelinks}

\usepackage[dvipsnames]{xcolor}

\usepackage{cancel}

\begin{document}

\title{Quantum correlations curvature, memory functions, and fundamental  bounds}

\author{Alexander Kruchkov}

\email{alexkruchkov@princeton.edu}

\affiliation{Department of Physics, Princeton University, Princeton, New Jersey 08544, USA}

\date {\today}

\begin{abstract}
We investigate fundamental bounds on the curvature of quantum correlation functions in imaginary time. Focusing first on topological phases, we show that quantum geometry can qualitatively modify the imaginary-time decay of correlations, leading to nontrivial curvature behavior beyond simple exponential scaling. More generally, we show a universal bound on correlation curvature that holds for interacting systems in thermal equilibrium, and establish  connection to leading invariants of the memory-function formalism. Our results identify imaginary-time curvature as a robust probe of intrinsic quantum timescales. 
\end{abstract}

\maketitle

The conquest for robust platforms for quantum information processing has renewed interest in quantum materials whose low-energy physics is shaped by topology and strong correlations \cite{Nayak2008}. Flat-band moiré materials, represented by twisted transition-metal dichalcogenides,  have emerged as leading candidates for Fractional Chern insulators \cite{Bernevig2025}, offering controllable realizations of topological order, anyonic excitations, and strongly entangled  states key for topological quantum computation. These systems place quantum geometry,  quantum metric and topological invariants, at the center of both equilibrium and ultrafast quantum phenomena \cite{Bernevig2025, Yu2025}.

Quantum correlations in imaginary time in such systems  present an interesting topic of investigation from both theoretical perspective and and the  possibility of quantum  measurement.  
Imaginary-time formulations play a central role in modern studies of quantum many-body systems because they convert oscillatory real-time dynamics into exponentially decaying evolution, producing smooth and well-conditioned correlation functions on the compact thermal circle $[0,\beta]$. This transformation eliminates  long-time instabilities, enabling direct and numerically controlled access to experimental observables such as susceptibilities, conductivities, quantum noise, and various type of quantum correlators. As a result, imaginary time underpins several of the most powerful nonperturbative numerical approaches to strongly correlated matter. In particular, quantum Monte Carlo methods are formulated natively in imaginary time \cite{Reynolds1990, Blankenbecler1981, Gull2011}. Different tensor-network techniques, including purification schemes and minimally entangled typical thermal states (METTS), similarly exploit imaginary-time evolution to construct thermal density matrices and equilibrium correlators in one and quasi-two dimensions \cite{Verstraete2004, White2009, Grundner2024, Phien2015, Czarnik2016}. Imaginary-time framework have also entered quantum simulation: variational imaginary-time evolution and related algorithms enable the preparation of ground and thermal states on noisy quantum hardware, bypassing the need for long coherent real-time dynamics \cite{McArdle2019, Motta2020, Sun2021, Turro2022, Nishi2023, Zhang2024}. 

Beyond numerical convenience, imaginary-time quantum correlators satisfy exact sum rules, providing a framework in which fundamental bounds on quantum dynamics can be derived directly from the first principles \cite{KRYU2023, Ji2025, Onishi2024}.
 Recent work has demonstrated that sum rules for quantum noise correlators are sensitive to the underlying quantum geometry \cite{Kruchkov2024}. Similar quantum-geometric mechanisms enable topological control of quantum speed limits\cite{Kruchkov2025}, and sets bounds on entanglement entropy  \cite{Paul2024, Tam2024, Kruchkov2025}. 
 Complementing these developments, ultrafast and nonlinear optical responses have emerged as powerful probes of quantum geometry, particularly in two-dimensional topological materials where quantum-geometric effects dominate ultrafast probes \cite{Yu2025}.

In this Letter, we identify the curvature of imaginary-time quantum correlations $\rho_0$ as a fundamental diagnostic of quantum dynamics,
\begin{align}
\rho_0 \equiv  \frac{ \sum_{i \omega_n}	(-1)^{n+1}  \omega_n^2 \, S_{ij} (i \omega_n) }{\sum_{i \omega_n}	S_{ij} (i \omega_n) } ,
\end{align}
where $S$ is a quantum noise correlator \cite{Kruchkov2024}.  Building on our previous work \cite{Kruchkov2023,Kruchkov2024,Kruchkov2025}, we investigate this curvature in topological systems and show that, in the presence of  nontrivial quantum metric, it acquires nontrivial characteristics [Eq.~\eqref{main}]. We further establish a connection between the extremal quantum correlation curvature and the coefficients of memory function, , highlighting the special role of thermal circle midpoint and Matsubara resummation with alternating signs.

\textit{Formalism.} 
The subject of this study is the \textit{nested commutator}
\begin{align}
[\mathcal H, [\mathcal H , O (\tau)] ]	,
\end{align}
which encodes information on the quantum correlation curvature in imaginary time $\tau$, 
\begin{align}
\rho (\tau) = \frac {	\langle [\mathcal H, [\mathcal H , \mathcal O (\tau)] ] \mathcal O \rangle }{\langle \mathcal O^2 \rangle }
\end{align}
($\mathcal H$ stands for the $\mathcal H$amiltonian).  As we discuss lateron,  curvature $\rho$ relates to the first nontrivial Liouvillian moment of the operator $\mathcal O$ and appears naturally as the leading coefficient of the memory functions  \cite{Mori1965}.
Building on Refs.~\cite{Kruchkov2024, Kruchkov2023,KRYU2023, Kruchkov2025}, we specialize on  current-current  imaginary-time correlator
\begin{align}
S_{ij} (\tau) = \langle J_i(\tau) J_j(0) \rangle .
\end{align}
We begin with the equal-time correlator \begin{align}
S_{ij} (0) \equiv \lim_{\tau \to 0} 
\langle J_i(\tau) J_j(0) \rangle 	, 
\label{noise}
\end{align}
which admits the sum-rule representation \cite{Kruchkov2024}
\begin{align}
S_{ij} (0)  = \frac{1}{\beta} \sum_{i \omega_n}  	S_{ij} ( i \omega_n) ,
\label{noise sum}
\end{align}
where $S_{ij} ( i \omega_n)$ is the Matsubara transform defined as 
\begin{align}
S_{ij} ( i \omega_n) = \int \limits_{0}^{\beta} d \tau e^{i \omega_n \tau} S (\tau). 
\end{align}	
Without loss of generality we consider longitudinal response (xx); the extension to the Hall channel is straightforward. 
 The quantity \eqref{noise sum} provides a finite and well-defined measure of quantum noise. At zero temperature it reduces to a geometric sum rule \cite{Kruchkov2024},
\begin{align}
S_{xx} (0) = \sum_{\vec k} \Delta ^2 (\vec k) \mathcal G_{xx} (\vec k)
\label{noise-sum}
\end{align}
where $\mathcal G_{xx}(\vec k)$  the quantum metric. 
As the quantum metric captures the gauge-invariant spread of electronic wave functions \cite{Marzari1997}, this contribution is manifestly nonnegative and finite \cite{Kruchkov2024}.

While the sum rule \eqref{noise sum} involves strictly positive contributions, additional structure emerges by considering the \textit{alternating Matsubara series} 
\begin{align}
 \sum_{i \omega_n}	(-1)^n S_{ij} (\omega) .
 \label{alternating}
\end{align}
Since the original series \eqref{noise} are convergent, the alternativ series are also convergent. However, while series \eqref{noise} are non-negative (there is always quantum noise in the system), the alternating series can sum up to zero. The physical importance of it is the indicator how the quantum correlator vanishes at the "furthest" imaginary time,
\begin{align}
 \sum_{i \omega_n}	(-1)^n S_{ij} (i \omega_n)
 \equiv \beta S \left(\tau = \nicefrac{\beta}{2} \right) .
 \label{alternating2}
\end{align}
Because in equilibrium imaginary time is compactified  onto a thermal circle of circumference $\beta$, the correlator satisfies $S(\tau)=S(\tau-\beta)$. This periodicity enforces an extremum at the antipodal point $\tau=\beta/2$, where the correlator acquires a well-defined curvature.  We quantify the curvature at $\tau=\beta/2$ through the series
\begin{align}
\rho_0 \equiv  \frac{ \sum_{i \omega_n}	(-1)^{n+1}  \omega_n^2 \, S_{ij} (i \omega_n) }{\sum_{i \omega_n}	S_{ij} (i \omega_n) } ,
\end{align}
where the denominator provides a natural normalization by the total quantum noise. As we show below, $\rho_0$ encodes intrinsic dynamical timescales and is subject to control through quantum metric.

\textit{Quantum correlation curvature in quantum-geometric systems.} We briefly recall the essential elements of the calculation before turning to the evaluation of the correlation curvature in the topological regime. Our analysis is carried out within the finite-temperature Matsubara formalism, ensuring causality and thermal equilibrium throughout. For concreteness, we focus on a two-dimensional lattice system; the arguments and results, however, extend straightforwardly to a broad class of quantum materials. The gauge-invariant lattice current operator, which serves as the central object in what follows, is defined as \cite{McKay2024}
\begin{align}
J_i (\vec q) = e \sum_{\vec k} \sum_{nm} \int d \alpha 
\langle u_{n, \vec k- \overline \alpha \vec q} | \mathcal V^i_{\vec k} | u_{m, \vec k + \alpha \vec q} \rangle 
\nonumber 
\\
\times c^{\dag}_{n, \vec k - \overline \alpha \vec q} c_{m \vec k+ \alpha \vec q}	,
\end{align}
where $c^{\dag}_{n \vec k}, c_{n \vec k}$
 are creating and annihilation operator in Bloch state $u_{n \vec k}$  with band index $n$ and quasimomentum $\vec k$, and $\mathcal V^i_{\vec k} = \partial_{k_i} \mathcal H _\vec k $, (we set $\hbar = k_B = e= 1$ and restore them in the end of the calculation). While retaining the full $\vec q$-dependence of the current operator is essential in certain contexts, here we focus on a broad class of systems for which this dependence is not crucial. Accordingly, we take the uniform limit $\vec q \to 0$ at the outset,
  \begin{align}
\vec J =  \sum_{\vec k} c^{\dag}_{\vec k} \boldsymbol {\mathcal V}_{\vec k }  c^{}_{\vec k}. 	
 \end{align}
With this definition of the current operator, the application of Wick’s theorem yields the current–current correlator (see Refs.~\cite{Kruchkov2024,Kruchkov2023}).
\begin{align}
S_{xx} (\tau) = 	\sum_{\vec k} \text{Tr} G_{\vec k} (- \tau) \mathcal V_x G_{\vec k} ( \tau) \mathcal V_x, 
\end{align}
or, upon proceeding to Matsubara transform $G(\tau) = \frac{1} {\beta} \sum_{i \omega'_n} e^{- i \omega'_n \tau } G(i \omega'_n)$, we obtain 

\begin{align}
S_{xx}(i\omega_0)
=
\frac{1}{\beta}
\sum_{\vec{k}}
\sum_{i\omega_n}
\operatorname{Tr}
\left[
G_{\vec{k}}(i\omega_n)
\mathcal V^x_{\vec k }
G_{\vec{k}}(i\omega_n + i\omega_0)
\mathcal V^x_{\vec k }
\right] ,
\end{align}
(here $i \omega_0$ is bosonic).  In what follows further, we consider a two-dimensional topological insulator with dispersive bands, separated by bandwidth $\Delta (\vec k)$. 
In topological phase, current operator acquires off-diagonal elements (we consider Fermi level in the gap); in the band basis $n,m$ one has \cite{Kruchkov2023,Blount} 
\begin{align}
\mathcal V_{nm}	 = \Delta_{mn} (\vec k) \langle u_{n \vec k} | \partial_{\vec k} u_{m \vec k} \rangle .
\end{align}

We find that the noise sum rule \eqref{noise-sum} gives 
 \begin{align}
 S_{xx} (0)  =  
\sum_{\vec k} \frac{\Delta^2 (\vec k)} {\tanh \frac{\beta \Delta (\vec k)}{2}} \mathcal G_{xx} (\vec k) .
\end{align}
At $T=0$, this reduces to Eq. \eqref{noise-sum}. 
Simultaneously, we  find that Eq. \ref{alternating2} gives
\begin{align}
\frac{d^2  S_{xx} \left(\nicefrac{\beta}{2} \right)  }{d \tau^2}  =  
\sum_{\vec k} \frac{\Delta^4 (\vec k)} {\sinh \frac{\beta \Delta (\vec k)}{2}} \mathcal G_{xx} (\vec k) ,
\end{align}
where $\mathcal G_{xx} (\vec k)$ is quantum metric of the lowest band, $\mathcal G_{ij} (\vec k) = \text{Re} \mathfrak G_{ij}(\vec k) $, 
\begin{align}
\mathfrak G_{ij} {\vec k}	= \langle  \partial_i u_{n \vec k} | 1- | u_{n \vec k} \rangle \langle |u_{n \vec k} | \partial_j u_{n \vec k}\rangle ,
\end{align}
where $\partial_i  = \frac{\partial }{\partial k_i}$. We thus obtain curvature 
\begin{align}
\rho_0 = \frac{ 
\sum_{\vec k} \frac{\Delta^4 (\vec k)} {\sinh \frac{\beta \Delta (\vec k)}{2}} \mathcal G_{xx} (\vec k)
}
{ \sum_{\vec k} \frac{\Delta^2 (\vec k)} {\tanh \frac{\beta \Delta (\vec k)}{2}} \mathcal G_{xx} (\vec k) } .
\label{main}
\end{align}
 For systems with a uniform or weakly varying quantum metric, this structure leads to an exponentially suppressed curvature as a function of inverse temperature $\beta$. In topological materials, however, the quantum metric is intrinsically nonuniform and strongly momentum dependent; its fluctuations enter the Brillouin-zone summation and preclude a simple  scaling. Moreover, the analysis above applies to noninteracting or weakly interacting limits, whereas interactions are an intrinsic feature of realistic quantum materials. These considerations naturally raise a broader question: does a model-independent, fundamental bound exist that constrains imaginary-time correlation curvature solely from first principles?

\textit{Bounds on quantum correlations curvature.} The equal-time correlator $S_{xx}(0)$ quantifies the intrinsic quantum noise of the system and provides a natural, but model-dependent—reference scale. To factor out this nonuniversal amplitude, we are interested in the normalized imaginary-time correlator,
\begin{align}
s_{xx} (\tau) \equiv \frac{ S_{xx} (\tau) }{ S_{xx}(0)} ,
\end{align}
We highlight again that this normalization is always well-defined since the denominator, encoding the measure of the noise in the system, is always nonzero in quantum systems \cite{Kruchkov2024}. 
 The full imaginary-time dependence of $s_{xx}(\tau)$ is fixed by linear response theory through its relation to experimentally accessible transport coefficients, in particular the longitudinal conductivity $\sigma_{xx}(\omega)$ \cite{Lederer2020},
\begin{align}
S_{xx} (\tau) = 	 \int \frac{d \omega}{2 \pi}   
\frac{\omega \cosh \left[ \frac{\beta \omega}{2}  - \tau \omega\right]}{\sinh \left[\frac{\beta \omega}{2} \right] } \text{Re} \sigma_{xx}(\omega) .
\end{align}
Furth3ron, the numerator reads 

 \begin{align}
\frac{d^2  S_{xx} (\tau)  }{d \tau^2}  = 	 \int \frac{d \omega}{2 \pi}   
\frac{\omega^3 \cosh \left[ \frac{\beta \omega}{2}  - \tau \omega\right]}{\sinh \left[\frac{\beta \omega}{2} \right] } \text{Re} \sigma_{xx}(\omega)
\label{curvature}
\end{align}
 while the denominator is 
 \begin{align}
S_{xx} (0) = 	 \int \frac{d \omega}{2 \pi}   
\frac{\omega }{\tanh \left[\frac{\beta \omega}{2} \right] } \text{Re} \sigma_{xx}(\omega)
\end{align}
This expression may be viewed as a special case of frequency sum rule for the optical conductivity \cite{KRYU2023}.

 \begin{figure}[b]
\includegraphics[width = 0.6 \columnwidth ]{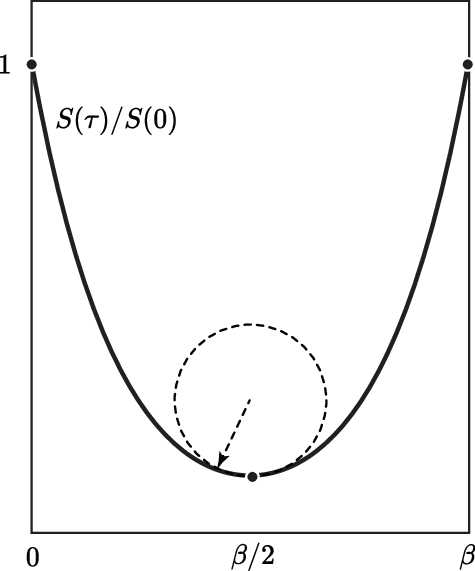}	
\caption{Curvature of quantum correlators at imaginary time $\tau = \nicefrac{\beta}{2}$ contains critical information on the internal timescales (see main text). Solid line: Imaginary-time correlator in a model of dispersionless weakly- interacting two-band Chern insulator ( \cite{Kruchkov2024,Kruchkov2023}).  We here consider two flat bands of Chern number $C=|1|$ separated by a topological gap $\Delta$, at temperature $T=0.2,\Delta$.  }
\end{figure} 

The curvature of the quantum correlator attains an extremum at $\tau=\beta/2$, a distinguished point on the thermal circle that we investigate in detail,
 \begin{align}
\frac{d^2  S_{xx} \left(\nicefrac{\beta}{2} \right)  }{d \tau^2}  = 	 \int \frac{d \omega}{2 \pi}   
\frac{\omega^3 }{\sinh \left[\frac{\beta \omega}{2} \right] } \text{Re} \sigma_{xx}(\omega) .
\end{align}
We therefore come to equality on the normalized correlator,
\begin{align}
\rho_0  =   \frac{\int \frac{d \omega}{2 \pi}   
\frac{\omega^3 }{\sinh \left[\frac{\beta \omega}{2} \right] } \text{Re} \sigma_{xx}(\omega) }
{ \int \frac{d \omega}{2 \pi}   
\frac{\omega }{\tanh \left[\frac{\beta \omega}{2} \right] } \text{Re} \sigma_{xx}(\omega)} .
\end{align}
We can further make change of variables to $\tilde \omega = \beta \omega /2$. We have  $\omega = 2 \tilde \omega /\beta = 2 T \tilde \omega$, 
 \begin{align}
\rho_0  = 	 \frac{4}{\beta^2}  \frac{\int \frac{d \tilde  \omega}{2 \pi}   
\frac{\tilde  \omega^3 }{\sinh \tilde \omega } \text{Re} \sigma_{xx}(2 \tilde \omega /\beta) }
{ \int \frac{d \tilde  \omega}{2 \pi}   
\frac{\tilde  \omega }{\tanh \tilde \omega } \text{Re} \sigma_{xx}(2 \tilde \omega /\beta)} .
\end{align}
For any physical system, both the numerator and denominator are convergent, signaling finite intrinsic correlation timescales. The denominator represents the total quantum noise and is therefore strictly nonzero. In addition, causality and thermal equilibrium imply $\mathrm{Re},\sigma_{xx}(\omega)\ge0$. These properties permit a direct comparison of the integrands, yielding a universal supremum bound,
\begin{align}
\rho_0 \leq	 \frac{4}{\beta^2}  
\text{sup} \frac{\tilde \omega^2}{\cosh \tilde \omega},
\label{supremum}
\end{align}
This bound is entirely independent of microscopic details of the system and holds whenever causality, Hermiticity, and boundary conditions on the thermal circle  are satisfied. The supremum on the right-hand side of Eq.~\eqref{supremum} occurs at $\tilde\omega_\ast$ determined by $\tilde\omega_\ast \tanh \tilde\omega_\ast = 2$, yielding $\tilde\omega_\ast \approx 2.07$ and $\sup\left(\tilde\omega^2/\cosh\tilde\omega\right)\approx 1.06$. We thus arrive at a universal upper bound on the curvature of quantum correlations,
 \begin{align}
\rho_0  \le 	 \frac{A^2}{\beta^2} ,
\ \ \  \ \ A \simeq 2.
\end{align} 
It is notable that the constant $A\sim 1$ is dimensionless, universal, and independent of microscopic details of the model. Its derivation relies only on three fundamental principles: \textit{Hermiticity}, \textit{causality}, and \textit{thermal equilibrium}.

 We note that higher-order nested correlators of the form
\begin{align}
 \langle \mathcal L^n \mathcal O (\tau) \mathcal O \rangle  \equiv \langle \underbrace{	\langle [\mathcal H, [\mathcal H , [\mathcal H ... [\mathcal H ,\mathcal O (\tau)] ]}_{\text{$n$ times}} \mathcal O \rangle
	\end{align} 
naturally arise in the memory-function representation of dynamical correlators and transport coefficients \cite{Mori1965}.  A straightforward calculation similar to \eqref{supremum} shows that higher-order nested commutators $\langle 	\langle [\mathcal H, [\mathcal H , [\mathcal H ... [\mathcal H ,  J (\tau) ] ]  J \rangle |_{\beta/2}$ are bounded by $1/\beta^n$  with prefactors that are independent of microscopic details. This structure highlights the special role of $\beta/2$ in constraining the hierarchy of dynamical moments.

\textit{Memory functions  and special role of thermal midpoint}.  
We now clarify the connection between the quantum correlation curvature at $\tau=\beta/2$ and experimentally accessible response functions. To this end, we introduce the Kubo–Mori inner product \cite{Mori1965},
\begin{align}
\left ( A | B  \right)  \equiv  \frac{1}{\beta}  \int_{0}^{\beta} \langle A^{\dag} (0) B (\tau) \rangle	
\end{align}
and invoke the memory-function formalism developed by Mori and collaborators \cite{Mori1965,Gotze1972}. Within this framework, the dynamics of the current operator are encoded in the current–current resolvent \cite{Gagliano1987,Dagotto1994,Georges1996,Foley2024} 
\footnote{In the absence of time-reversal symmetry additional coefficients appear in the continued-fraction representation; the discussion remains unchanged.}\begin{align}
\mathcal S ({z}) = (J |(z- \mathcal L)^{-1}|J) 
= \cfrac{(J|J)}{
z - \cfrac{b_1^2}{
z - \cfrac{b_2^2}{
z - \cfrac{b_3^2}{\ddots}
}}}	
\label{Mori}
\end{align}
The continued fraction in \eqref{Mori} contains coefficients
\begin{align}
b_1^2 = \frac{( J| \mathcal L^2 |  J)}{(J|J)}	, 
\ \ \ \ \ 
b_2^2 = \frac{( J| \mathcal L^4 |  J)}{(J| L^2 | J)}  - 	\frac{( J| \mathcal L^2 |  J)}{(J|J)}	, 
\end{align}
et cetera.  The optical conductivity is obtained, up to the diamagnetic contribution, from the analytically continued (retarded) resolvent as
$
\sigma (\omega ) = {\mathcal S^R (\omega) }/{i \omega }	
$. 
This representation permits a direct estimate on the leading coefficient, defining the denominator of resolvent as  \begin{align}
b_1^2 = \frac{\int_{0}^{\beta} \langle J (0)  [H, [H, J (\tau)]] \rangle}{\int_{0}^{\beta} \langle J (0) J (\tau) \rangle}
\geq
\frac{\langle J (0)  [H, [H, J (\tau)]] \rangle|_{\tau = \beta/2}}{ \langle J (0) J (\tau) \rangle_{\tau = 0}}
\end{align}
Consequently, $b_1^{-1}$ defines an intrinsic microscopic timescale  for current operator dynamics; it is the timescale of initial decay of the memory kernel. In quantum-geometric systems, this timescale follows nontrivial scaling \eqref{main}.

\

\textit{Discussion.} The results of this work are twofold.
First, at the level of explicit formulas, we derive closed expressions for the imaginary-time curvature of current–current correlations in topological systems [Eq.~\eqref{main}]. We show that, in the presence of nontrivial quantum geometry, this curvature need not exhibit the exponential suppression characteristic of topologically trivial phases. Instead, \textit{fluctuations of the quantum metric} qualitatively modify the long-imaginary-time behavior of correlations, providing a direct geometric contribution to their curvature. More broadly, we establish a general framework in which the curvature of quantum correlations at the \textit{midpoint of the thermal circle}, $\tau=\beta/2$, encodes intrinsic dynamical information. The special role of this imaginary time is not incidental: it follows directly from the memory-function formalism and from the structure of the leading Krylov coefficients defined via the Kubo–Mori inner product. Because the corresponding imaginary-time kernels are extremized at $\tau=\beta/2$, the curvature at this point naturally defines an intrinsic microscopic timescale of the system. In this sense, $\tau=\beta/2$ emerges as a distinguished locus where equilibrium correlations are maximally sensitive to operator dynamics.
 Taken together, these results suggest that imaginary-time curvature, particularly at $\beta/2$, provides a robust  probe of quantum dynamics, complementing frequency-resolved and real-time transport and spectral diagnostics.

 We expect this perspective to be broadly applicable, including in strongly correlated and topological materials, and to motivate further investigation of the special role of thermal circle midpoint quantum dynamics, and \textit{Matsubara resummation with alternating signs}.

 \bibliography{Refs}

\end{document}